\documentclass{article}
\usepackage{arxiv}
\usepackage[utf8]{inputenc}
\usepackage[T1]{fontenc}
\usepackage{hyperref}
\usepackage{url}
\usepackage{booktabs}
\usepackage{amsfonts}
\usepackage{nicefrac}
\usepackage{microtype}
\usepackage{lipsum}
\usepackage{graphicx}
\usepackage{multirow}
\usepackage[thinlines]{easytable}
\usepackage[table,xcdraw]{xcolor}
\usepackage{float}
\usepackage{mdframed}
\usepackage[export]{adjustbox}
\usepackage{scrextend}
\usepackage[table,xcdraw]{xcolor}
\usepackage{longtable}
\usepackage{lscape}
\usepackage{booktabs}
\usepackage{caption}

\graphicspath{ {./images/} }
\title{Predictive analysis of Bitcoin price \\ considering social sentiments.}

\author{
	Pratikkumar Prajapati \\
	\small{\textit{Department of Computer Science, San Jos\'e State University}} \\
	\small{\textit{San Jos\'e, California, USA}} \\
	\small{\textit{\href{mailto:pratikkumar.prajapati@sjsu.edu}{pratikkumar.prajapati@sjsu.edu}}}
}

\begin{document}
	\maketitle
	\begin{abstract}
		We report on the use of sentiment analysis on news and social media to analyze and predict the price of Bitcoin. Bitcoin is the leading cryptocurrency and has the highest market capitalization among digital currencies. Predicting Bitcoin values may help understand and predict potential market movement and future growth of the technology. Unlike (mostly) repeating phenomena like weather, cryptocurrency values do not follow a repeating pattern and mere past value of Bitcoin does not reveal any secret of future Bitcoin value. Humans follow general sentiments and technical analysis to invest in the market. Hence considering people’s sentiment can give a good degree of prediction. We focus on using social sentiment as a feature to predict future Bitcoin value, and in particular, consider Google News and Reddit posts. We find that social sentiment gives a good estimate of how future Bitcoin values may move. We achieve the lowest test RMSE of 434.87 using an LSTM that takes as inputs the historical price of various cryptocurrencies, the sentiment of news articles and the sentiment of Reddit posts. \\
	\end{abstract}


	\section{Introduction}
	Bitcoin \cite{bitcoin} has sparked a gigantic interest in cryptocurrency and blockchain technology. Since the inception of Bitcoin, cryptocurrency has gained the trust of the general population. Bitcoin has achieved the highest market capitalization among all of the cryptocurrencies. As of this writing Bitcoin market capitalization is more than 134 billion US dollars \cite{coinmarketcap}. Bitcoin gains this market value as there a huge demand for this cryptocurrency. The demand for the cryptocurrency directly translates into people’s trust in the Bitcoin and the underneath technology. Since people’s trust is involved in the rise of the cryptocurrency market, the sentiment of the general population does make a huge impact on the future of the cryptocurrency market capitalization. Hence, we use the sentiment of people in an attempt to predict future Bitcoin prices. https://news.google.com (Google News) is a nice source for collecting news posted by various journalists around the globe. Google News also provides the capability to search the news based on selected keywords and its search tools also have a feature of crawling the news based on the date of the news release. While Google News gives opinions of the various journalists we also focus on sentiments of the general population. https://www.Reddit.com (Reddit) is also one of the most famous social platforms where people can post anonymously. We also consider the sentiment of messages posted on Reddit to predict the Bitcoin price movement. Along with sentiments, we have included historical price and volume of Litecoin \cite{litecoin} and Ethereum \cite{ethereum}. We have trained various machine learning models to learn about the correlation between all these features and results are analyzed.

	\section{Related work}
	Predicting Bitcoin price has been the interest of many researchers recently. \cite{arxiv191111819} has used Support Vector Machines to predict the Bitcoin price and discussed trading strategies. \cite{arxiv190901268} has performed a comprehensive analysis using various machine learning models by feeding historical price data. \cite{sdS0264999317304558} analyzed the correlation of Bitcoin price with its volume movement. Other research \cite{transgraph} was done to predict price of Bitcoin using its transaction graph. We have not found any research using various social sentiments and so our research is focused on that novel aspect.

	\section{Data engineering}
	
	There were five major steps performed to create the dataset. Extracting data from (1) Google News, (2) Reddit, and (3) cryptocurrency exchanges were core steps in collecting the data. In the fourth step, the text data from Google News and Reddit posts were used to perform sentiment analysis. In the fifth step, sentiment results were combined with cryptocurrency historical price to build the final dataset. All data were collected from January 1, 2018, to November 20, 2019, with a time interval of one hour. The only exception to this is about Google News, whose data were collected per day basis.
	
	\subsection{Google News data}
	
	\subsubsection{Collection}
	A Python-based crawling script,  google\_news\_scraper.py \footnote{Sample implementation can be found at  \url{https://github.com/pratikpv/predicting_bitcoin_market}. See section 'Reference Implementation' for details.}, is written using supplement libraries to get data from Google News. The script uses ‘requests’ \cite{requestspy} Python library to post HTTP GET requests with specially formatted URL. The API supported by Google News has a feature to search for news by given keywords and a date range. See Fig. \ref{fig:fig1} for the URL that Google News expects.

	\begin{figure}[htb!]
		\centering
		\begin{mdframed}
			https://www.google.com/search?q=\textit{bitcoin+cryptocurrency}\&hl=en\&gl=us\&as\_drrb=b\\
			\&tbas=0\&tbs=cdr:1,cd\_min:\textit{{min\_date}},cd\_max:\textit{{max\_date}}
		\end{mdframed}
		\caption{Google News URL template to extract news. “bitcoin+cryptocurrency“ is the keyword passed, min\_date and max\_date is the date range of interest.}
		\label{fig:fig1}
		
	\end{figure}

	The crawling script posts the HTTP GET requests for each date and parses the results. The results include hyperlinks to actual news articles and the first few lines of the news content. In order to perform sentiment analysis as accurately as possible, we obtained the whole news article. The ‘newspaper’ \cite{newspaperpy} Python library is used to download the news articles whose URLs are found from initial request to Google News. The first ten news articles are downloaded for each day and a Pandas data-frame \cite{pandaspy} is created for the day. In the end, the data-frame is saved as a CSV file on disk or further processing later. See Table. \ref{tab:gog-news-table} for the sample data collected by the crawling script.
	
	\begin{longtable}[c]{@{}lllll@{}}
		\caption{Data collected by the Google News crawling script. Each row represents the news posted on a given date. Each
			column represents the news text. Top 9 news posts are collected for each day.}
		\label{tab:gog-news-table}\\
		\toprule
		\textbf{date} & \textbf{news\_1\_text} & \textbf{news\_2\_text} & \textbf{...} & \textbf{news\_9\_text} \\* \midrule
		\endfirsthead
		\endhead
		\bottomrule
		\endfoot
		\endlastfoot
		2018-01-01 & First it was the stock mar... & Japan’s GDP Grows Due to B... & ... & NaN \\
		2018-01-02 & Bitcoin's dominance of the... & With 2017 now in the books... & ... & Decrypting cryptocurrency:... \\
		2018-01-03 & (Reuters) - Bitcoin was th... & Ripple is soaring, but you... & ... & FanDuel is jumping on theb... \\
		2018-01-04 & Robinhood Expands Across t... & Mr. McCaleb has since crea... & ... & One of Egypt’s most promin... \\
		2018-01-05 & This Week in Cryptocurrenc... & Ripple is making waves. Th... & ... & The word blockchain is oft... \\
		... & ... & ... & ... & ... \\* \bottomrule
	\end{longtable}

	\subsubsection{Sentiment Analysis}
	The text of the news article is passed to pre-trained sentiment analyzer machine learning models and numerical values are extracted from the model. All the news articles of the day are analyzed this way and then the final value of the sentiment is divided by the number of articles parsed for the day to get the normalized sentiment value of the day. 
	Flair \cite{flair}, Valence Aware Dictionary and sEntiment Reasoner (VADER) \cite{vader}, and TextBlob \cite{textblob} models are used to perform sentiment analysis. Flair is a state-of-the-art natural language processing (NLP) model and allows named entity recognition (NER), part-of-speech tagging (PoS), sense disambiguation and classification. VADER is a lexicon and rule-based sentiment analysis tool that is specifically attuned to sentiments expressed in social media, so VADER would also be suitable for our purpose. TextBlob provides simple capabilities for NLP tasks such as PoS tagging, noun phrase extraction, sentiment analysis, classification and more. We have chosen three different sentiment analysis models to get different views and perspectives of the media news and posts.
	google\_news\_sentiment\_analysis.py script is written to perform the sentiment analysis. See Table. \ref{tab:gog-senti-data} for the data generated by the sentiment analysis script. Eventually, we need each row to represent data of a one-hour span to match with other tables. For Google News, there is no reliable method available to get the news posted on an hourly basis. So, we have decided to replicate each day's reports twenty-four times to match with the hourly format.
	
	\begin{longtable}[c]{@{}llllllll@{}}
		\caption{Data generated by the Google News sentiment analysis script.}
		\label{tab:gog-senti-data}\\
		\toprule
		\textbf{date} & \textbf{\begin{tabular}[c]{@{}l@{}}gnews\_\\ flair\end{tabular}} & \textbf{\begin{tabular}[c]{@{}l@{}}gnews\_tb\_\\ polarity\end{tabular}} & \textbf{\begin{tabular}[c]{@{}l@{}}gnews\_tb\_\\ subjectivity\end{tabular}} & \textbf{\begin{tabular}[c]{@{}l@{}}gnews\_sid\_\\ pos\end{tabular}} & \textbf{\begin{tabular}[c]{@{}l@{}}gnews\_sid\_\\ neg\end{tabular}} & \textbf{\begin{tabular}[c]{@{}l@{}}gnews\_sid\_\\ neu\end{tabular}} & \textbf{\begin{tabular}[c]{@{}l@{}}gnews\_sid\_\\ com\end{tabular}} \\* \midrule
		\endfirsthead
		\endhead
		\bottomrule
		\endfoot
		\endlastfoot
		2018-01-01 & 0.0426 & 0.0678 & 0.3232 & 0.0651 & 0.0252 & 0.9096 & 0.6247 \\
		2018-01-02 & 0.5485 & 0.0675 & 0.3065 & 0.1091 & 0.0397 & 0.8510 & 0.6256 \\
		2018-01-03 & -0.0764 & 0.0917 & 0.3232 & 0.1137 & 0.0293 & 0.8571 & 0.6213 \\
		2018-01-04 & 0.3247 & 0.0457 & 0.3838 & 0.0966 & 0.0477 & 0.8560 & 0.2898 \\
		2018-01-05 & -0.1257 & 0.1146 & 0.4000 & 0.0890 & 0.0367 & 0.8742 & 0.9214 \\
		... & ... & ... & ... & ... & ... & ... & ... \\* \bottomrule
	\end{longtable}

	\subsection{Reddit messages data}
	
	\subsubsection{Collection}
	A Python-based crawling script, called download\_data\_from\_reddit.py, is written to get data from Reddit. The script uses Pushshift APIs \cite{pushshift}. One of the advantages of this method is that it does not need API secret keys from Reddit and there is no limit on data or number posts to request (as of this writing). The script searched in rBitcoin subreddit with keyword ‘Bitcoin’ to search the posts. See Table. \ref{tab:redd-post-data}. For sample data generated by this script.

	\begin{longtable}[c]{@{}llllllllll@{}}
		\caption{Data generated by the Reddit crawling script. [post\_id, title, selftext, url, author, score, publish\_date, num\_of\_comments, permalink, flair] attributes of each post are collected.}
		\label{tab:redd-post-data}\\
		\toprule
		\textbf{post\_id} & \textbf{title} & \textbf{selftext} & \textbf{url} & \textbf{author} & \textbf{score} & \textbf{publish\_date} & \textbf{\begin{tabular}[c]{@{}l@{}}num\_\\ of\_\\ comments\end{tabular}} & \textbf{...} & \textbf{flair} \\* \midrule
		\endfirsthead
		\endhead
		\bottomrule
		\endfoot
		\endlastfoot
		7ne3y9 & If Governments t... & Many state govts... & ... & bit... & 18 & \begin{tabular}[c]{@{}l@{}}2018-01-01\\     00:10:48\end{tabular} & 14 & ... & NaN \\
		7ne5h2 & Inactive bitcoin... & So I have some b... & ... & day... & 1 & \begin{tabular}[c]{@{}l@{}}2018-01-01\\     00:22:47\end{tabular} & 3 & ... & NaN \\
		7ne5in & Unconfirmed tran... & hey guys, happy ... & ... & The... & 15 & \begin{tabular}[c]{@{}l@{}}2018-01-01 \\      00:23:08\end{tabular} & 7 & ... & NaN \\
		7ne60x & Bitcoin taxes & So if I received... & ... & fail... & 5 & \begin{tabular}[c]{@{}l@{}}2018-01-01\\      00:27:51\end{tabular} & 35 & ... & NaN \\
		7ne6dg & 2018 we getting ... & NaN & ... & Pal... & 0 & \begin{tabular}[c]{@{}l@{}}2018-01-01\\      00:30:58\end{tabular} & 13 & ... & \begin{tabular}[c]{@{}l@{}}low\\ quality\end{tabular} \\
		... & ... & ... & ... & ... & ... & ... & ... & ... & ... \\* \bottomrule
	\end{longtable}
	
	\subsubsection{Sentiment Analysis}
	Sentiment analysis is performed on Title and self-text fields of the posts using reddit\_post\_sentiment\_analysis.py script. Like Google News sentiment analysis Flair, VADER, and TextBlob models are used to perform sentiment analysis. See Table. \ref{tab:redd-senti-data}. for the data generated by the script.
	
	\begin{longtable}[c]{@{}llllllll@{}}
		\caption{Data generated by the Reddit posts sentiment analysis script.}
		\label{tab:redd-senti-data}\\
		\toprule
		\textbf{timestamp} & \textbf{reddit\_flair} & \textbf{\begin{tabular}[c]{@{}l@{}}reddit\_tb\_\\ polarity\end{tabular}} & \textbf{\begin{tabular}[c]{@{}l@{}}reddit\_tb\_\\ subjectivity\end{tabular}} & \textbf{\begin{tabular}[c]{@{}l@{}}reddit\_\\ sid\_pos\end{tabular}} & \textbf{\begin{tabular}[c]{@{}l@{}}reddit\_\\ sid\_neg\end{tabular}} & \textbf{\begin{tabular}[c]{@{}l@{}}reddit\_\\ sid\_neu\end{tabular}} & \textbf{\begin{tabular}[c]{@{}l@{}}reddit\_\\ sid\_com\end{tabular}} \\* \midrule
		\endfirsthead
		\endhead
		\bottomrule
		\endfoot
		\endlastfoot
		2018-01-01 00:10:48 & -0.9971 & 0.1100 & 0.3350 & 0.077 & 0.066 & 0.857 & 0.5672 \\
		2018-01-01 00:22:47 & -0.9999 & 0.0222 & 0.2667 & 0.085 & 0.102 & 0.814 & 0.2846 \\
		2018-01-01 00:23:08 & -0.9991 & 0.0179 & 0.4227 & 0.174 & 0.129 & 0.697 & 0.6944 \\
		2018-01-01 00:27:51 & -0.9909 & 0.0000 & 0.0000 & 0.090 & 0.074 & 0.835 & 0.0900 \\
		2018-01-01 00:30:58 & 0.9731 & 0.3750 & 0.7500 & 0.419 & 0.000 & 0.581 & 0.5574 \\
		... & ... & ... & ... & ... & ... & ... & ... \\* \bottomrule
	\end{longtable}

	At this point, one more step is needed to streamline this table. These posts are bucketized on hour basis and all respective sentiment values are normalized by its arithmetic mean. Reddit sentiment analyzer script does this transformation and the final table of data is shown in Table. \ref{tab:redd-senti-buck-data}.
	
	\begin{longtable}[c]{@{}llllllll@{}}
		\caption{Reddit data bucketized by hour basis.}
		\label{tab:redd-senti-buck-data}\\
		\toprule
		\textbf{timestamp} & \textbf{reddit\_flair} & \textbf{\begin{tabular}[c]{@{}l@{}}reddit\_tb\_\\ polarity\end{tabular}} & \textbf{\begin{tabular}[c]{@{}l@{}}reddit\_tb\_\\ subjectivity\end{tabular}} & \textbf{\begin{tabular}[c]{@{}l@{}}reddit\_\\ sid\_pos\end{tabular}} & \textbf{\begin{tabular}[c]{@{}l@{}}reddit\_\\ sid\_neg\end{tabular}} & \textbf{\begin{tabular}[c]{@{}l@{}}reddit\_\\ sid\_neu\end{tabular}} & \textbf{\begin{tabular}[c]{@{}l@{}}reddit\_\\ sid\_com\end{tabular}} \\* \midrule
		\endfirsthead
		\endhead
		\bottomrule
		\endfoot
		\endlastfoot
		2018-01-01 00:00:00 & 0.1876 & -0.0778 & 0.1444 & 0.0000 & 0.1575 & 0.8425 & -0.2484 \\
		2018-01-01 01:00:00 & -0.2672 & 0.0462 & 0.2579 & 0.1056 & 0.0919 & 0.8025 & 0.1813 \\
		2018-01-01 02:00:00 & -0.3008 & -0.0772 & 0.2732 & 0.0320 & 0.0193 & 0.9487 & 0.3301 \\
		2018-01-01 03:00:00 & 0.0825 & 0.2425 & 0.4044 & 0.1563 & 0.0250 & 0.8187 & 0.2801 \\
		2018-01-01 04:00:00 & 0.4437 & 0.2353 & 0.3214 & 0.1531 & 0.0051 & 0.8417 & 0.3418 \\
		... & ... & ... & ... & ... & ... & ... & ... \\* \bottomrule
	\end{longtable}
	
	\subsection{Cryptocurrency data}
	
	\subsubsection{Collection and merging}
	Bitcoin, Litecoin, and Ethereum data are also used to train the model. Initial data are collected from crypto-data-download web \cite{crypto-data-download}, but it was missing some recent data. Hence, download\_data\_from\_binance.py script was written to get the most recent data. Binance \cite{binance} has nice and easy to use Python-based client APIs. Data collected by this script is shown in Table. \ref{tab:crypto-data}. This script retains ['open', 'high', 'low', 'close', 'volume'] values of Bitcoin and ['close', 'volume'] values of other currencies.
	
	% \usepackage{booktabs}
	% \usepackage{longtable}
	% Note: It may be necessary to compile the document several times to get a multi-page table to line up properly
	\begin{longtable}[c]{@{}lllllllll@{}}
		\caption{Data generated by the cryptocurrency data downloader script.}
		\label{tab:crypto-data}\\
		\toprule
		\textbf{timestamp} & \textbf{\begin{tabular}[c]{@{}l@{}}open\_\\ BTCUSDT\end{tabular}} & \textbf{...} & \textbf{\begin{tabular}[c]{@{}l@{}}close\_\\ BTCUSDT\end{tabular}} & \textbf{\begin{tabular}[c]{@{}l@{}}volume\_\\ BTCUSDT\end{tabular}} & \textbf{\begin{tabular}[c]{@{}l@{}}close\_\\ LTCUSD\end{tabular}} & \textbf{\begin{tabular}[c]{@{}l@{}}volume\_\\ LTCUSD\end{tabular}} & \textbf{\begin{tabular}[c]{@{}l@{}}close\_\\ ETHUSD\end{tabular}} & \textbf{\begin{tabular}[c]{@{}l@{}}volume\_\\ ETHUSD\end{tabular}} \\* \midrule
		\endfirsthead
		\endhead
		\bottomrule
		\endfoot
		\endlastfoot
		\begin{tabular}[c]{@{}l@{}}2018-01-01\\     00:00:00\end{tabular} & 13820.26 & ... & 13557.88 & 210.21 & 222.24 & 590.23 & 728.77 & 625.29 \\
		\begin{tabular}[c]{@{}l@{}}2018-01-01\\     01:00:00\end{tabular} & 13557.88 & ... & 13262.85 & 191.93 & 215.20 & 698.36 & 724.27 & 710.89 \\
		\begin{tabular}[c]{@{}l@{}}2018-01-01\\     02:00:00\end{tabular} & 13262.85 & ... & 13320.00 & 169.46 & 215.36 & 464.55 & 722.11 & 849.26 \\
		\begin{tabular}[c]{@{}l@{}}2018-01-01\\     03:00:00\end{tabular} & 13320.00 & ... & 13372.00 & 80.46 & 219.30 & 407.81 & 733.19 & 556.40 \\
		... & ... & ... & ... & ... & ... & ... & ... & ... \\* \bottomrule
	\end{longtable}
	
	\subsection{Combining all data}
	To feed data into the machine learning models, all data from Google News, Reddit posts and cryptocurrency data are merged together using merge\_data\_files.py script. The data generated by this script is shown in Table. \ref{tab:final-data}

	\begin{longtable}[c]{@{}llllllllll@{}}
		\caption{Final data generated by combining all the tables. All these columns {[}open\_BTCUSDT, high\_BTCUSDT, low\_BTCUSDT, close\_BTCUSDT, volume\_BTCUSDT, close\_LTCUSD, volume\_LTCUSD, close\_ETHUSD, volume\_ETHUSD, gnews\_flair, gnews\_tb\_polarity, gnews\_tb\_subjectivity, gnews\_sid\_pos, gnews\_sid\_neg, gnews\_sid\_neu, gnews\_sid\_com, reddit\_flair, reddit\_tb\_polarity, reddit\_tb\_subjectivity, reddit\_sid\_pos, reddit\_sid\_neg, reddit\_sid\_neu, reddit\_sid\_com{]} are merged for final set of data.}
		\label{tab:final-data}\\
		\toprule
		\textbf{timestamp} & \textbf{\begin{tabular}[c]{@{}l@{}}open\_\\ BTCUSDT\end{tabular}} & \textbf{...} & \textbf{\begin{tabular}[c]{@{}l@{}}volume\_\\ ETHUSD\end{tabular}} & \textbf{\begin{tabular}[c]{@{}l@{}}gnews\_\\ flair\end{tabular}} & \textbf{...} & \textbf{\begin{tabular}[c]{@{}l@{}}gnews\_\\ sid\_com\end{tabular}} & \textbf{\begin{tabular}[c]{@{}l@{}}reddit\_\\ flair\end{tabular}} & \textbf{...} & \textbf{\begin{tabular}[c]{@{}l@{}}reddit\_\\ sid\_com\end{tabular}} \\* \midrule
		\endfirsthead
		\endhead
		\bottomrule
		\endfoot
		\endlastfoot
		2018-01-01 00:00:00 & 13820.26 & ... & 728.77 & 0.0426 & ... & 0.6247 & 0.0000 & ... & 0.0000 \\
		2018-01-01 01:00:00 & 13557.88 & ... & 724.27 & 0.0426 & ... & 0.6247 & -0.2672 & ... & 0.1813 \\
		2018-01-01 02:00:00 & 13262.85 & ... & 722.11 & 0.0426 & ... & 0.6247 & -0.3008 & ... & 0.3301 \\
		2018-01-01 03:00:00 & 13320.00 & ... & 733.19 & 0.0426 & ... & 0.6247 & 0.0825 & ... & 0.2801 \\
		2018-01-01 04:00:00 & 13372.00 & ... & 738.59 & 0.0426 & ... & 0.6247 & 0.4437 & ... & 0.3418 \\
		... & ... & ... & ... & ... & ... & ... & ... & ... & ... \\* \bottomrule
	\end{longtable}
	
	\section{Training the machine learning models}
	
\begin{figure}[htb!]
		\centering
		\includegraphics[width=0.7\textwidth, height=6cm]{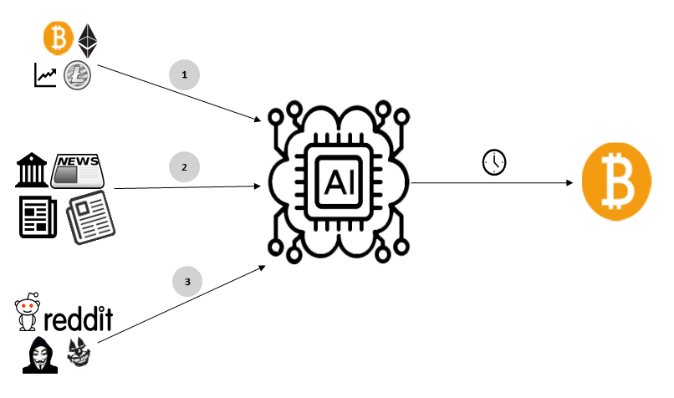}
		\caption{High-level architecture of the model. (1) Historical data of Bitcoin, Litecoin, and Ethereum (2) Sentiment of news media (3) Sentiment of Reddit messages are passed to Artificial Intelligence models and Bitcoin values are predicted.}
		\label{fig:model_arch}
	\end{figure}

	Fig. \ref{fig:model_arch} shows the graphical view of the data passed to the machine learning model. With the recent advancement of Graphics Processing Units (GPUs) based systems, training on the huge amount of data using artificial intelligence has become feasible.
	
	In this research project, we use various artificial neural networks such as Long Short-Term Memory (LSTM) \cite{lstm}, Gated Recurrent Unit (GRU) \cite{gru}, one-dimensional convolution neural network (1D-CNN) \cite{cnn} to predict the price of Bitcoin. All these models are implemented using Keras \cite{keras}. 
	\\ \\
	Unlike feedforward networks, LSTM has the capability to get feedback from past data and process the sequence of data using its input, output and forget gates with internal cells. Since LSTM can act on a time-series of data or sequence of data, it can produce good accuracy in predictions in the given scenario. GRU is similar to LSTM but with lesser hyper-parameters and no output gate. We choose GRU as well because it needs lesser training and might help in cases the LSTM-based model could overfit. We have also experimented with 1D-CNN and LSTM stacking to evaluate if 1D-CNN can identify some features in local sequence and then LSTM can look over the longer sequence of data. As the data is in time-series format these models should be suitable for the experiments. We are passing the historical price of Bitcoin, Litecoin and Ethereum, Google News sentiment reports and Reddit posts sentiment reports to the various models. 
	\\ \\ 
	Seventeen experiments are performed and the accuracy of the models is calculated using Root Mean Squared Error (RMSE). 
	
	\begin{equation}
	RMSE = \sqrt{\frac {\sum_{i=1}^{n} (Predicted_i - Actual_i)^2 }{n}}
	\end{equation}
	
	RMSE is chosen as it penalizes more on large errors and it increases with the variance of the frequency distribution of error magnitudes. Results are discussed in later sections.
	
	\section{Reference implementation}
	
	All code written for the research has been uploaded to \url{https://www.github.com} and the following subsections give details on the  repositories.
	
	\subsection{Google News data}
	
	\begin{itemize}
		\item	google\_news\_scraper.py: Google News crawling script.
		\item	google\_news\_sentiment\_analysis.py: Google News sentiment analyzer script.
	\end{itemize}

	Both scripts are posted at: 
	\url{https://github.com/pratikpv/google_news_scraper_and_sentiment_analyzer}
	
	\subsection{Reddit message data}
	
	\begin{itemize}
		\item	download\_data\_from\_reddit.py: Reddit messages crawling script.
		\item	reddit\_post\_sentiment\_analysis.py: Reddit messages sentiment analyzer script.
	\end{itemize}	
	Both scripts are posted at: \url{https://github.com/pratikpv/reddit_scraper_and_sentiment_analyzer}
	
	\subsection{Cryptocurrency data}
	
	\begin{itemize}
		\item	download\_data\_from\_binance.py: Client script to download cryptocurrency data from Binance exchange.
	\end{itemize}
	
	The script is posted at: \url{https://github.com/pratikpv/cryptocurrency_data_downloader}
	
	\subsection{Data merging}
	
	\begin{itemize}
		\item merge\_data\_files.py: Script to combine all of the data.
	\end{itemize}
	The script is posted at: \url{https://github.com/pratikpv/predicting_bitcoin_market}
	
	\subsection{Experiments}
	Jupyter notebooks of all the experiments are posted at:
	\url{https://github.com/pratikpv/predicting_bitcoin_market/tree/master/experiments}

	\section{Results and analysis}
	A total of seventeen experiments are performed with various hyperparameters of the models and various configurations of the models. Tables \ref{tab:results-tab1}, \ref{tab:results-tab2}, and \ref{tab:results-tab3} give the details on the features, parameters and results aspects of all the experiments.  Notes in tables represent the high-level architecture of the model used. See Table \ref{tab:lagend-notes} for description of notes. \\ \\	
	Open, high, low, close value, and volume features are used for Bitcoin (BTC). Closing value and volume features are used for Litecoin (LTC) and Ethereum (ETH). Flair, Textblob polarity, Textblob subjectivity, VADER positive, and VADER negative features are used for Google News and Reddit text data. ‘x’ marked cells in the tables represent that the feature is selected for the respective experiment. \\ \\	
	Lookback days in the model parameters section represent the sliding window of the LSTM and GRU based model. E.g. 60 lookback days mean we are looking back up to 60 days.  Each row in the data represents one-hour span of the data, so 60 days of lookback would select a total of 60 * 24 records. Various configurations of layers are also tried. Batch-size and epochs are fixed at 128 and 5 respectively, as any change in these parameters were either not improving the results or were overfitting the models.
	RMSE is used as the loss function to optimize the model. Mean Absolute Error (MAE) is also calculated at the end for completeness, but MAE was not used as the loss function.
	
	\setlength{\tabcolsep}{0.5em} % for the horizontal padding
	{\renewcommand{\arraystretch}{1.7}% for the vertical padding
		\begin{table}[h!]
			\centering
			\caption{Notes legend}
			\label{tab:lagend-notes}
			\resizebox{\textwidth}{!}{%
				\begin{tabular}{|c|l|}
					\hline
					\textbf{Note \#} & \multicolumn{1}{c|}{\textbf{Remarks}}                                                                                                             \\ \hline
					1                & Sentiment values as the sum of Reddit and Google News. E.g. Flair value of Google News and Reddit posts are added and passed as a single feature. \\ \hline
					2                & LSTM model                                                                                                                                        \\ \hline
					3                & LSTM $\rightarrow$ LSTM stacked model                                                                                                             \\ \hline
					4                & LSTM $\rightarrow$ GRU stacked model                                                                                                              \\ \hline
					5                & GRU model                                                                                                                                         \\ \hline
					6                & GRU $\rightarrow$ GRU stacked model                                                                                                               \\ \hline
					7                & LSTM $\rightarrow$ GRU stacked model                                                                                                              \\ \hline
					8                & Conv1D $\rightarrow$ LSTM stacked model                                                                                                           \\ \hline
				\end{tabular}%
			}
		\end{table}
	}
	
	\setlength{\tabcolsep}{1em} % for the horizontal padding
	{\renewcommand{\arraystretch}{1.6}% for the vertical padding
		\begin{table}[H]
			\centering
			\caption{Results of experiments 1 to 6}
			\label{tab:results-tab1}
			\begin{tabular}{|c|c|c|c|c|c|c|c|}
				\hline
				\multicolumn{2}{|c|}{\textbf{Contents}}                  & \textbf{Expr 1} & \textbf{Expr 2} & \textbf{Expr 3} & \textbf{Expr 4} & \textbf{Expr 5} & \textbf{Expr 6} \\ \hline
				\multirow{19}{*}{Features}    & open\_BTCUSDT            &                 &                 &                 & x               & x               & x               \\ \cline{2-8} 
				& high\_BTCUSDT            &                 &                 &                 & x               & x               & x               \\ \cline{2-8} 
				& low\_BTCUSDT             &                 &                 &                 & x               & x               & x               \\ \cline{2-8} 
				& close\_BTCUSDT           & x               & x               & x               & x               & x               & x               \\ \cline{2-8} 
				& volume\_BTCUSDT          & x               & x               & x               & x               & x               & x               \\ \cline{2-8} 
				& close\_LTCUSD            & x               & x               & x               & x               &                 & x               \\ \cline{2-8} 
				& volume\_LTCUSD           & x               & x               & x               & x               &                 & x               \\ \cline{2-8} 
				& close\_ETHUSD            & x               & x               & x               & x               &                 & x               \\ \cline{2-8} 
				& volume\_ETHUSD           & x               & x               & x               & x               &                 & x               \\ \cline{2-8} 
				& gnews\_flair             & x               & x               & x               & x               &                 &                 \\ \cline{2-8} 
				& gnews\_tb\_polarity      & x               & x               & x               & x               &                 &                 \\ \cline{2-8} 
				& gnews\_tb\_subjectivity  & x               & x               & x               & x               &                 &                 \\ \cline{2-8} 
				& gnews\_sid\_pos          & x               & x               & x               & x               &                 &                 \\ \cline{2-8} 
				& gnews\_sid\_neg          & x               & x               & x               & x               &                 &                 \\ \cline{2-8} 
				& reddit\_flair            & x               & x               & x               & x               &                 &                 \\ \cline{2-8} 
				& reddit\_tb\_polarity     & x               & x               & x               & x               &                 &                 \\ \cline{2-8} 
				& reddit\_tb\_subjectivity & x               & x               & x               & x               &                 &                 \\ \cline{2-8} 
				& reddit\_sid\_pos         & x               & x               & x               & x               &                 &                 \\ \cline{2-8} 
				& reddit\_sid\_neg         & x               & x               & x               & x               &                 &                 \\ \hline
				\multirow{4}{*}{Model Params} & look back days           & 60              & 60              & 120             & 120             & 60              & 60              \\ \cline{2-8} 
				& layers                   & 32              & 64              & 64              & 64              & 32              & 32              \\ \cline{2-8} 
				& batch size               & 128             & 128             & 128             & 128             & 128             & 128             \\ \cline{2-8} 
				& epochs                   & 5               & 5               & 5               & 5               & 5               & 5               \\ \hline
				\multirow{4}{*}{Results}      & Train RMSE               & 769.11          & 829.19          & 1254.8          & 231.42          & 154.11          & 642.9           \\ \cline{2-8} 
				& Test RMSE                & 2490.4          & 2632.1          & 3541.8          & 434.87          & 173.72          & 556.5           \\ \cline{2-8} 
				& Train MAE                & 572.49          & 679.46          & 1027.4          & 181.29          & 82.72           & 510.85          \\ \cline{2-8} 
				& Test MAE                 & 2454.6          & 2597.7          & 3530.9          & 421.56          & 116.36          & 477.92          \\ \hline
				\multicolumn{2}{|c|}{Notes}                              & 1,2             & 1,2             & 1,2             & 1,2             & 2               & 2               \\ \hline
			\end{tabular}
		\end{table}
		}
		
	\setlength{\tabcolsep}{1em} % for the horizontal padding
	{\renewcommand{\arraystretch}{1.6}% for the vertical padding
		\begin{table}[H]
			\centering
			\caption{Results of experiments 7 to 12}
			\label{tab:results-tab2}
			\begin{tabular}{|c|c|c|c|c|c|c|c|}
				\hline
				\multicolumn{2}{|c|}{\textbf{Contents}}                  & \textbf{Expr 7} & \textbf{Expr 8} & \textbf{Expr 9} & \textbf{Expr10} & \textbf{Expr11} & \textbf{Expr12} \\ \hline
				\multirow{19}{*}{Features}    & open\_BTCUSDT            &                 &                 &                 &                 &                 & x               \\ \cline{2-8} 
				& high\_BTCUSDT            &                 &                 &                 &                 &                 & x               \\ \cline{2-8} 
				& low\_BTCUSDT             &                 &                 &                 &                 &                 & x               \\ \cline{2-8} 
				& close\_BTCUSDT           & x               & x               & x               & x               & x               & x               \\ \cline{2-8} 
				& volume\_BTCUSDT          & x               & x               & x               & x               & x               & x               \\ \cline{2-8} 
				& close\_LTCUSD            & x               & x               & x               & x               & x               & x               \\ \cline{2-8} 
				& volume\_LTCUSD           & x               & x               & x               & x               & x               & x               \\ \cline{2-8} 
				& close\_ETHUSD            & x               & x               & x               & x               & x               & x               \\ \cline{2-8} 
				& volume\_ETHUSD           & x               & x               & x               & x               & x               & x               \\ \cline{2-8} 
				& gnews\_flair             & x               &                 & x               &                 &                 & x               \\ \cline{2-8} 
				& gnews\_tb\_polarity      & x               &                 &                 & x               &                 & x               \\ \cline{2-8} 
				& gnews\_tb\_subjectivity  & x               &                 &                 & x               &                 & x               \\ \cline{2-8} 
				& gnews\_sid\_pos          & x               &                 &                 &                 & x               & x               \\ \cline{2-8} 
				& gnews\_sid\_neg          & x               &                 &                 &                 & x               & x               \\ \cline{2-8} 
				& reddit\_flair            &                 & x               & x               &                 &                 & x               \\ \cline{2-8} 
				& reddit\_tb\_polarity     &                 & x               &                 & x               &                 & x               \\ \cline{2-8} 
				& reddit\_tb\_subjectivity &                 & x               &                 & x               &                 & x               \\ \cline{2-8} 
				& reddit\_sid\_pos         &                 & x               &                 &                 & x               & x               \\ \cline{2-8} 
				& reddit\_sid\_neg         &                 & x               &                 &                 & x               & x               \\ \hline
				\multirow{4}{*}{Model Params} & look back days           & 60              & 60              & 60              & 60              & 60              & 120             \\ \cline{2-8} 
				& layers                   & 32              & 32              & 32              & 32              & 32              & 64              \\ \cline{2-8} 
				& batch size               & 128             & 128             & 128             & 128             & 128             & 128             \\ \cline{2-8} 
				& epochs                   & 5               & 5               & 5               & 5               & 5               & 5               \\ \hline
				\multirow{4}{*}{Results}      & Train RMSE               & 766.14          & 706.31          & 813.34          & 814.08          & 751.09          & 534.23          \\ \cline{2-8} 
				& Test RMSE                & 2406.1          & 1943.3          & 1746            & 2282.4          & 789.86          & 1514.8          \\ \cline{2-8} 
				& Train MAE                & 600.18          & 571.6           & 624.37          & 646.16          & 561.81          & 399.22          \\ \cline{2-8} 
				& Test MAE                 & 2369.2          & 1892.5          & 1703.6          & 2248.5          & 745.71          & 1507.4          \\ \hline
				\multicolumn{2}{|c|}{Notes}                                   & 2               & 2               & 1,2             & 1,2             & 1,2             & 1,3             \\ \hline
			\end{tabular}
		\end{table}
	}
	
	\setlength{\tabcolsep}{1em} % for the horizontal padding
	{\renewcommand{\arraystretch}{1.6}% for the vertical padding
		\begin{table}[H]
			\centering
			\caption{Results of experiments 13 to 17}
			\label{tab:results-tab3}
			\begin{tabular}{|c|c|c|c|c|c|c|}
				\hline
				\multicolumn{2}{|c|}{\textbf{Contents}}                  & \textbf{Expr13} & \textbf{Expr14} & \textbf{Expr15} & \textbf{Expr16} & \textbf{Expr17} \\ \hline
				\multirow{19}{*}{Features}    & open\_BTCUSDT            & x               & x               & x               &                 & x               \\ \cline{2-7} 
				& high\_BTCUSDT            & x               & x               & x               &                 & x               \\ \cline{2-7} 
				& low\_BTCUSDT             & x               & x               & x               &                 & x               \\ \cline{2-7} 
				& close\_BTCUSDT           & x               & x               & x               & x               & x               \\ \cline{2-7} 
				& volume\_BTCUSDT          & x               & x               & x               & x               & x               \\ \cline{2-7} 
				& close\_LTCUSD            & x               & x               & x               & x               & x               \\ \cline{2-7} 
				& volume\_LTCUSD           & x               & x               & x               & x               & x               \\ \cline{2-7} 
				& close\_ETHUSD            & x               & x               & x               & x               & x               \\ \cline{2-7} 
				& volume\_ETHUSD           & x               & x               & x               & x               & x               \\ \cline{2-7} 
				& gnews\_flair             & x               & x               & x               & x               & x               \\ \cline{2-7} 
				& gnews\_tb\_polarity      & x               & x               & x               &                 & x               \\ \cline{2-7} 
				& gnews\_tb\_subjectivity  & x               & x               & x               &                 & x               \\ \cline{2-7} 
				& gnews\_sid\_pos          & x               & x               & x               &                 & x               \\ \cline{2-7} 
				& gnews\_sid\_neg          & x               & x               & x               &                 & x               \\ \cline{2-7} 
				& reddit\_flair            & x               & x               & x               & x               & x               \\ \cline{2-7} 
				& reddit\_tb\_polarity     & x               & x               & x               &                 & x               \\ \cline{2-7} 
				& reddit\_tb\_subjectivity & x               & x               & x               &                 & x               \\ \cline{2-7} 
				& reddit\_sid\_pos         & x               & x               & x               &                 & x               \\ \cline{2-7} 
				& reddit\_sid\_neg         & x               & x               & x               &                 & x               \\ \hline
				\multirow{4}{*}{Model Params} & look back days           & 60              & 60              & 60              & 60              & 60              \\ \cline{2-7} 
				& layers                   & 32              & 32              & 32              & 32              & 32              \\ \cline{2-7} 
				& batch size               & 128             & 128             & 128             & 128             & 128             \\ \cline{2-7} 
				& epochs                   & 5               & 5               & 5               & 5               & 5               \\ \hline
				\multirow{4}{*}{Results}      & Train RMSE               & 377.4           & 440.89          & 433.69          & 788.32          & 1249.3          \\ \cline{2-7} 
				& Test RMSE                & 983.94          & 1314.9          & 1177            & 2146.1          & 2615.3          \\ \cline{2-7} 
				& Train MAE                & 270.37          & 350.08          & 329.24          & 614.04          & 1003.9          \\ \cline{2-7} 
				& Test MAE                 & 977.53          & 1308.5          & 1171            & 2137.6          & 2605            \\ \hline
				\multicolumn{2}{|c|}{Notes}                              & 1,4             & 1,5             & 1,6             & 1,7             & 1,8             \\ \hline
			\end{tabular}
		\end{table}
	}

	\section{Conclusion}
	
\begin{figure}[H]
		\centering
		\includegraphics[width=0.9\textwidth, height=7cm]{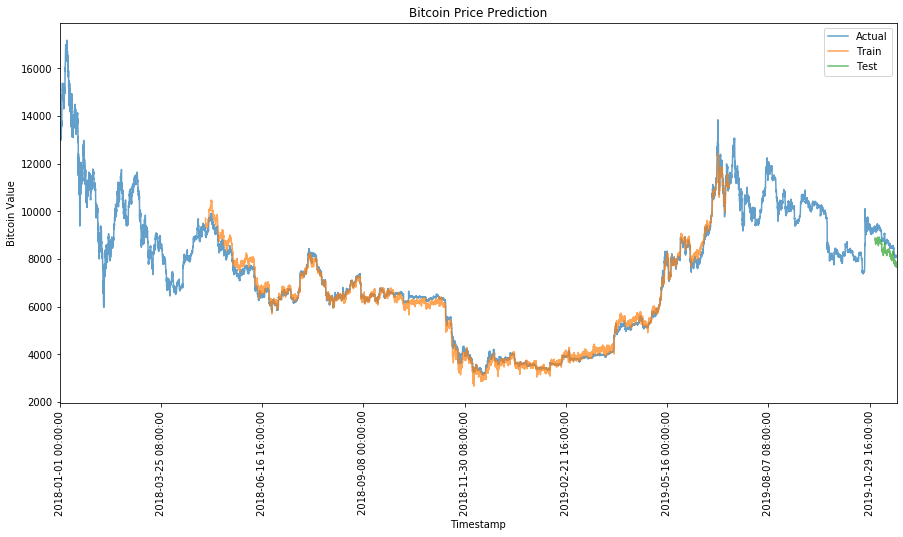}
		\caption{Data visualization for Experiment 4.}
		\label{fig:fig10}
	\end{figure}

\begin{figure}[H]
		\centering
		\includegraphics[width=0.9\textwidth, height=7cm]{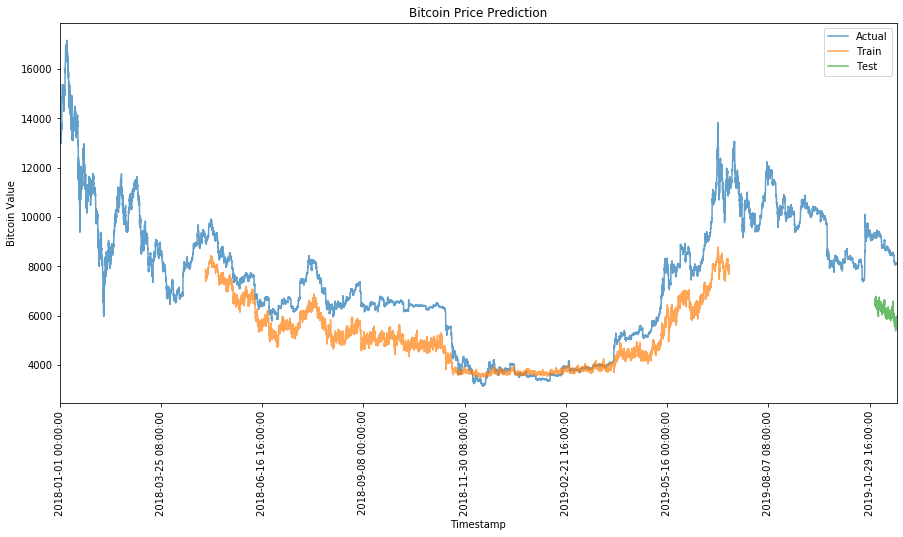}
		\caption{Data visualization for Experiment 17.}
		\label{fig:fig11}
	\end{figure}

	Experiment \#4 gives the lowest Test RMSE of 434.87 while using all the features, including sentiment analysis reports. Experiment \#5 is also interesting as even though it does not use any sentiment analysis, it gives the lowest RMSE of all. But intuitively this would not be much reliable as Bitcoin or any currency/stock market value does not depend merely on past values. Considering a longer history of data might prove this intuitive argument. Fig. \ref{fig:fig10} shows the results of experiment \#4 visually and is the best of all experiments while considering sentiment features. It can be seen that predicted test values (in green line) are following very closely to actual values of Bitcoin. Experiment \#17 gives the highest Test RMSE of 2615.3 and is the worst-performing model. Fig. \ref{fig:fig11} shows the results of experiment \#17 visually and it can be seen that the green line of prediction is not following the actual values closely. Experiment \#7 uses Google News data and cryptocurrency data, while experiment \#8 uses Reddit data and cryptocurrency data. Experiment \#8 gives lower RMSE compared to experiment \#7. Intuitively we also conclude that Reddit post sentiment values are more co-related then Google News sentiments.
				
	\section{Future work}
	
	More historical data can be collected using the scripts given in this research project and it would be interesting to prove Experiment 4 can perform better than Experiment 5. Also, more social sentiment can be collected from other platforms like Facebook, Twitter and more. Furthermore, it would be interesting to consider any correlation between cryptocurrency price movements and fluctuations in the world’s leading stock and foreign exchange markets.
	
	\section*{Acknowledgments}
	The author thanks Volkmar Frinken, Guha Jayachandran, and Shriphani Palakodety for lectures and guidance during the Blockchain Technologies course in Fall 2019.
	
	\bibliographystyle{unsrt}
	\bibliography{references}

\end{document}